\documentclass[aps,prb,twocolumn,nofootinbib,showpacs]{revtex4}
\usepackage{graphicx}
\usepackage{amsfonts}
\usepackage{amssymb}

\usepackage{graphicx}
\usepackage{amsfonts}
\usepackage{amssymb}
\usepackage{epsfig}
\usepackage{psfrag}
\usepackage{dsfont}
\usepackage{bm}
\newcommand{\beqn}{\begin{eqnarray}}
\newcommand{\eeqn}{\end{eqnarray}}
\newcommand{\be}{\begin{equation}}
\newcommand{\ee}{\end{equation}}

\def\beq{\begin{equation}}
\def\eeq{\end{equation}}
\def\beqn{\begin{eqnarray}}
\def\eeqn{\end{eqnarray}}

\def\s1{$s_{\alpha}$}
\def\s2{$s_{\gamma}$}
\def\s3{$s_{\delta}$}
\def\c1{$c_{\alpha}$}
\def\c2{$c_{\gamma}$}
\def\c3{$c_{\delta}$}
\def\s{Stueckelberg~}

\newcommand{\mathsym}[1]{{}}

\begin{document}
\begin{center}
\end{center}

\title{Resonant Inelastic X-ray Scattering (RIXS) Spectra for Ladder Cuprates }

\author{W. Al-Sawai,$^1$ R.S. Markiewicz,$^1$ M.Z. Hasan,$^2$ and A. Bansil$^1$ }

\affiliation{$^1$Department of Physics, Northeastern University,
 Boston, MA 02115, USA \\$^2$Department of Physics, Joseph Henry Laboratories of Physics, Princeton University, Princeton, NJ 08544,
 USA}
 \date{\today}

\begin{abstract}

The ladder compound Sr$_{14}$Cu$_{24}$O$_{41}$ is of interest both
as a quasi-one-dimensional analog of the superconducting cuprates
and as a superconductor in its own right when Sr is substituted by
Ca. In order to model resonant inelastic x-ray scattering (RIXS)
spectra for this compound, we investigate the simpler
SrCu$_{2}$O$_{3}$ system in which the crystal structure contains
very similar ladder planes. We approximate the LDA dispersion of
SrCu$_{2}$O$_{3}$ by a Cu only two-band tight-binding model. Strong
correlation effects are incorporated by assuming an
anti-ferromagnetic ground state. The available angle-resolved
photoemission (ARPES) and RIXS data on the ladder compound are found
to be in reasonable accord with our theoretical predictions.

\end{abstract}
\maketitle

\section{Introduction}

Resonant inelastic x-ray scattering (RIXS) is a second-order optical
process in which there is a coherent absorption and emission of
X-rays in resonance with electronic excitations.\cite{KoShin} RIXS
can probe charge excitations extending to fairly high energies of up
to $\sim$ 8 eV. This allows the analysis of electronic states over a
wide energy range, including electron correlation effects
originating from strong electron-electron Coulomb repulsion,
providing thus a powerful tool for investigating Mott physics in
solids.

The chain-ladder compound Sr$_{14}$Cu$_{24}$O$_{41}$ exhibits very
interesting magnetic, transport and properties. It has attracted
wide attention due to the discovery of a superconducting phase in
highly Ca-doped samples at high pressure \cite{nagata} and charge
order of the doped ladder \cite{abbamonte}. The compound possesses
an incommensurate layered structure consisting of alternating layers
of sublattices involving CuO$_{2}$ chains and Cu$_{2}$O$_{3}$
ladders. The superconductivity arises on the ladders, making them a
quasi-one-dimensional analog of the cuprates. Very recently, K-edge
RIXS data on the ladder compound has been
reported\cite{Hasan,Ishii}, providing motivation for undertaking
corresponding theoretical modeling of the spectra. Here, we attempt
to do so by considering the simpler analog compound
SrCu$_{2}$O$_{3}$. This should be a good approximation since
interlayer coupling in Sr$_{14}$Cu$_{24}$O$_{41}$ is negligible
\cite{arai}, and both compounds have very similar ladder planes with
similar hopping parameters \cite{muller}. Specifically, we obtain
K-edge RIXS spectra within a mean field approach for momentum
transfer along as well as perpendicular to the direction of the
ladders. A two-band Cu-only tight-binding model is used in which
strong correlation effects are incorporated by treating an
antiferromagnetic (AFM) ground state.
\section {Electronic Structure and the Two Band Model}

The spin-ladder compound SrCu$_{2}$O$_{3}$ possesses the
orthorhombic structure with space group Cmmm in which
Cu$_{2}$O$_{3}$ planes are stacked with Sr atoms sandwiched between
these planes.\cite{John} Fig. 1 shows the detailed arrangement of Cu
and O atoms in the Cu$_{2}$O$_{3}$ planes. This so-called `trellis
structure' involves Cu-O ladders where successive ladders are seen
to be offset by half a unit cell. We obtained the band structure of
SrCu$_{2}$O$_{3}$ self-consistently using a full-potential, all
electron scheme within the local density approximation
(LDA)\cite{wien2k,bansil99_1}. The first principles bands were
fitted by a 2-band tight-binding (TB) model in the vicinity of the
Fermi energy, and provided the basis for RIXS computations presented
in this study. Fig. 2 shows the first-principles as well as the TB
bands along several high symmetry lines in the Brillouin zone (BZ).
There are seen to be only two bands around the Fermi energy, which
display large dispersion along the ladder direction $\Gamma$-Z, and
a relatively smaller dispersion along the perpendicular $\Gamma$-X
direction. In the first-principles band structure, both these bands
are dominated by states of Cu $d_{x^2-y^2}$ character whose weight
is given by the color bar on the right hand side of
Fig. 2.\\
\begin{figure}[t]
\vspace*{.2in} \hspace*{-.2in} \centering
\includegraphics[width=8cm,height=10cm]{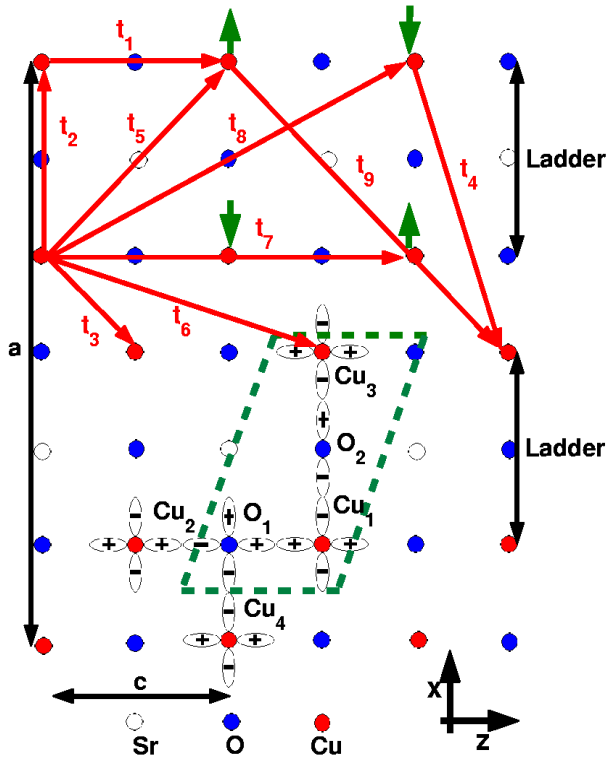}
\caption{ (color online) Arrangement of Cu and O atoms in the
Cu$_{2}$O$_{3}$ planes in the form of a series of ladders offset by
half a unit cell along the c-axis. Unfilled circles give the
location of out-of-plane Sr atoms. The green parallelepiped marks a
primitive unit cell around which the orientation of Cu
$d_{x^2-y^2}$, O ${p_x}$, and O ${p_y}$ orbitals is shown. Red
arrows give the specific hopping parameters used in two-band model
fits to the band structure. The AFM ordering of spins in the ladders
is depicted by the green arrows. } \label{fig1}
\end{figure}
\begin{figure}[t]
\vspace*{.2in} \hspace*{-.2in} \centering
\includegraphics[width=8cm,height=8cm]{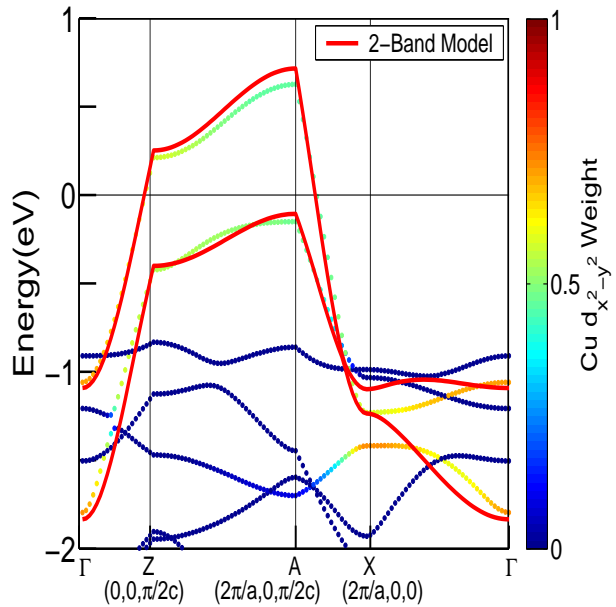}
\caption{(color online) First-principles LDA band structure of
SrCu$_{2}$O$_{3}$ along several high symmetry lines in the
irreducible Brillouin zone is shown by dots of various colors. The
colors of the dots code the weight of Cu $d_{x^2-y^2}$ character in
the associated wavefunctions as indicated by the color-bar on the
right hand side of the figure. Solid red lines give the 2-band
tight-binding model fit to the LDA bands near the Fermi energy. }
\label{figd1}
\end{figure}
In making a TB fit to the two aformentioned LDA bands near the Fermi
energy, we have adapted a Cu-only 2-band model suggested in
Ref.~\onlinecite{muller}. The detailed form of the TB Hamiltonian is
discussed in the Appendix. The TB bands are seen from Fig. 2 to
provide a good fit to the LDA bands near the Fermi energy.  Our
values of various parameters, i.e. the on-site energy $\epsilon_0$
and the hopping parameters $t_1-t_9$, are seen from Table 1 to be in
reasonable accord with those of Ref.~\onlinecite{muller}. The
meaning of specific overlap terms involved in defining $t_1-t_9$ is
clarified by the red arrows in Fig. ~1. The present TB model
includes not only the nearest neighbor hopping terms $t_1-t_3$, but
also the longer range hoppings $t_4-t_9$. Interestingly, we find
that the inter-ladder dispersion (i.e. along Z-A in Fig. 2) cannot
be fitted well using only a nearest-neighbor hopping model. Table 1
shows that the intra-ladder hopping parameters ($t_1$, $t_2$ and
$t_5$) are generally larger than the inter-ladder terms such as
$t_3$, $t_4$ and $t_6$. This can be understood with reference to
Fig. 1 where orientation of the Cu-$d_{x^2-y^2}$ and O-$p_x$ and
$p_y$ orbitals is sketched on a few sites. An intra-ladder Cu-O-Cu
path with a bond angle of $180^{o}$ (e.g. Cu$_{2}$-O$_{1}$-Cu$_{1}$)
will be expected to provide a larger orbital overlap than an
inter-ladder path with a $90^{o}$ bond (e.g.
Cu$_{4}$-O$_{1}$-Cu$_{1}$). The glide symmetry of SrCu$_{2}$O$_{3}$
leads to some dispersion anomalies, including extra degeneracies at
the zone boundaries and an apparent $4\pi$ periodicity of the
dispersion along the ladder. A $2\pi$ symmetry can be effectively
restored by including different cuts along $k_{x}$ as shown in Fig.
3(a). Similar anomalies in $c$-axis dispersion due to a glide
symmetry are also found in Bi$_2$Sr$_2$CaCu$_2$O$_8$
(Bi2212)\cite{Bansilkz}.
\begin{figure}[t]
\vspace*{.2in} \hspace*{-.2in} \centering
\includegraphics[width=8cm,height=7cm]{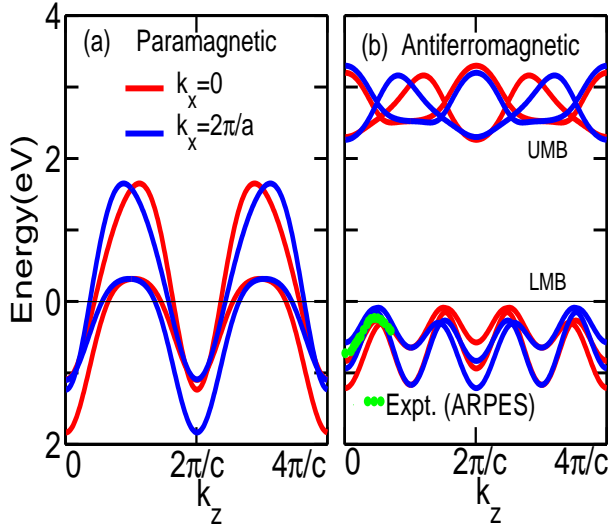}
\caption{(color online) (a) Paramagnetic, and (b) AFM dispersions
within the 2-band TB model as a function of $k_z$ at $k_x=0$ (red
lines) and $k_x=2\pi/a$ (blue lines). Experimental ARPES datapoints
are shown as green dots.\cite{takahashi} } \label{figdd2}
\end{figure}
\begin{table}[h]
\caption{\label{tab:table2}TB parameters for 2-band model.}
\begin{ruledtabular}
\begin{tabular}{clclcc}
Parameter         &This work&      &Ref.~\onlinecite{muller}&    \\
\hline $\epsilon_{0}$        &-0.0350(eV)& &-0.0450(eV)&        \\
$t_{1}$         &~0.5650&        &~0.5650&               \\
$t_{2}$         &~0.3800&       &~0.3950&               \\
$t_{3}$         &~0.0400&       &~0.0400&           \\
$t_{4}$         &~0.0520&        &~0.0500&            \\
$t_{5}$         &-0.1200&      &-0.1150&            \\
$t_{6}$         &~0.0700&        &~0.0400&            \\
$t_{7}$         &~0.0750&      &~0.0750&            \\
$t_{8}$         &~0.0057&       &~0.0050&            \\
$t_{9}$         &-0.0115&       &-0.0200&            \\
\end{tabular}
\end{ruledtabular}
 \end{table}
The trellis compound shows the presence of short-range spin order
with a spin gap which is consistent with theoretical predictions
\cite{dagotto}. Due to the $180^{o}$ Cu$_{1}$-O$_{1}$-Cu$_{2}$
bonds, the spins are strongly coupled antiferromagnetically on the
legs and the rungs of the ladders as indicated by green arrows in
Fig. 1.\cite{alain} However, the displacement of successive ladders
with respect to each other frustrates the development of long range
AFM order. One nevertheless expects the electronic system to
experience significant AFM fluctuations, which are presumably
sufficient to impose an underlying dispersion characteristic of the
AFM order. In this spirit, we have approximated the correlation
effects within a Hartree-Fock model of an itinerant AFM, as in the
planar cuprates\cite{Kusko}. Taking the on-site energy to be $U$=
3.3 eV $\sim $ 6$t$, the magnetization was computed
self-consistently to be $m$=0.43. The AFM Hamiltonian is given in
the Appendix and the resulting dispersions are shown in Fig.~3(b).
Comparison with the paramagnetic solution shows that a large gap of
$\sim$2.3~eV opens up between the upper (UMB) and the lower magnetic
bands (LMB). The theoretical LMBs display the characteristic
backfolding near $k_z=\pi /2c$, which is in accord with the
experimentally observed dispersion (green dots in Fig. 3(b) via
ARPES \cite{takahashi}, and is reminiscent of a similar effect in
the insulating planar cuprates.
 \begin{figure}[t]
\vspace*{.2in} \hspace*{-.2in} \centering
\includegraphics[width=7cm,height=7.1cm]{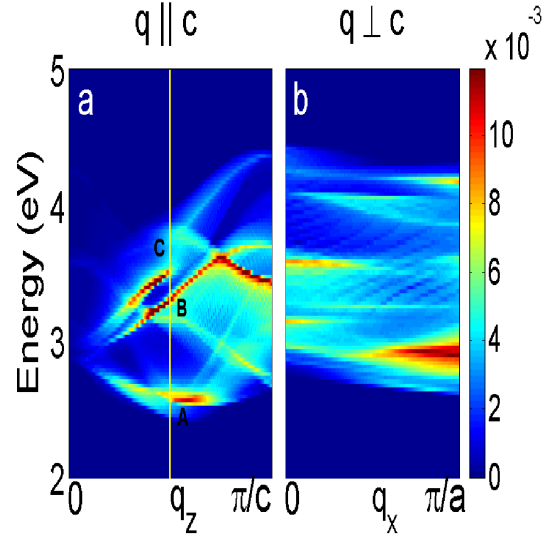}
\caption{(color online) Color plots show computed RIXS spectra from
SrCu$_{2}$O$_{3}$ for momentum transfer along (a): The ladder
direction $q_z$ and (b): The direction $q_x$ perpendicular to the
ladders. Theoretical spectra are not resolution broadened in order
to highlight their intrinsic structure. Yellow vertical line marks
the momentum transfer value where the detailed spectra, including
features marked A-C, are considered in Fig. 5. }
 \label{fig4}
\end{figure}
 \section {RIXS Spectra} Our computations of the the K-edge RIXS cross
section for the Cu
 $1s\rightarrow 4p$ core level excitation are based on the expression
\cite{Mark,nomura}

 \be
\begin{array}{c}
I(\textbf{q},\omega,\omega_i)=(2\pi)^3N|W(\omega,\omega_i)|^2\\\times\sum_{j,j^\prime,\textbf{k}
}\chi^{"}_{jj^\prime}(\textbf{q},\textbf{k},\omega)|M{ij}(\textbf{k})|^2

\end{array}
\ee where

\be
\begin{array}{c}
\chi^{"}_{jj^\prime}(\textbf{q},\textbf{k},\omega)=\delta(\omega+E_{j}
(\textbf{k})-E_{j^\prime}(\textbf{k}+\textbf{q}))\\
\times n_j(\textbf{k})[1-n_{j^{\prime}}(\textbf{q}+\textbf{k})],
\end {array}
\ee $n_j(k)$ is the electron occupation of the $j^{th}$ band and
$E_j(k)$ is the corresponding energy dispersion obtained by
self-consistently solving the two-band AFM Hamiltonian (see
Appendix), and

\be
\begin{array}{c}
W(\omega_f,\omega_i)=|\gamma|\Sigma_{k_1}\frac{V_d}{D(\omega_{i},\textbf{k}_1)D(\omega_{f},\textbf{k}_2)}.
\end{array}
\ee Here,
$D(\omega,\textbf{k})=\omega+\varepsilon_{1s}-\varepsilon_{4p}(k_1)+i\Gamma_{1s}$,
$\gamma$ is the matrix element for scattering from $1s$ to $4p$, and
$V_d$ is the core-hole potential in $3d$ level. $\omega_i(\omega_f)$
and $q_i(q_f)$ denote the initial (final) energy and momentum,
respectively, of the photon, and $\omega=\omega_i-\omega_f$ and
$q=q_i-q_f$ give the energy and momentum transferred in the
scattering process. Since Cu ${1s}$ is a core state, the associated
energy band $\varepsilon_{1s}(k)$ is assumed dispersionless. The Cu
${4p}$ band dispersion $ \varepsilon_{4p}(k)$ is modeled by a 2D-TB
model with nearest neighbor hopping. $\Gamma_{1s}$ is the decay rate
of core hole taken to be 0.8 eV. The matrix element $M_{i,j}$
associated with the interaction between the core hole and $3d$
levels around the Fermi energy is

\begin{figure}[t]
\vspace*{.2in} \hspace*{-.2in} \centering
\includegraphics[width=8.2cm,height=7cm]{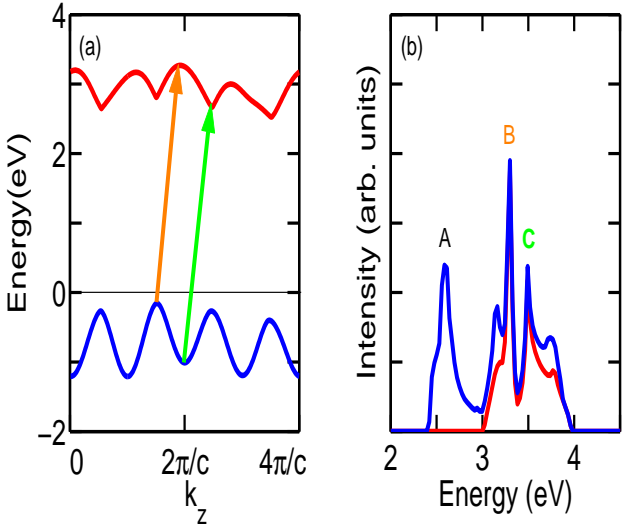}
\caption{(color online) (a) Lowest and uppermost tight-binding AFM
bands in SrCu$_{2}$O$_{3}$ at $k_x=0$. (b) Red line gives partial
contribution to the RIXS spectrum at $q_z=\pi/{2c}$ (marked by
yellow vertical line in Fig. 4(a)) from transitions between the two
bands in (a). Blue line gives the corresponding total RIXS spectrum.
Features A-C are discussed in the text. } \label{fig5}
\end{figure}

\be
 \begin{array}{c}
M_{ij}(\textbf{k})=\sum_{l,\sigma,\sigma^\prime}e^{i\textbf{q}\cdot
\textbf{R}_{l}}\alpha_{l}X^j_{l\sigma}(\textbf{k})\Lambda^{j}_{\sigma,
\sigma^{\prime}}(\omega,\textbf{q})X^{j^{\prime}}_{l\sigma^{\prime}}(\textbf{k}+\textbf{q}),
\end{array}
\ee in terms of the eigenvectors $X^j_{l\sigma}$ of the AFM
Hamiltonian, where $\sigma$ denotes electron spin and $l$ an orbital
index. $\alpha_l\equiv V_l/V_d$, where $V_l$ is the Coulomb
interaction between a core hole and an electron on atom $l$
separated by a distance $R_l$. Here we approximate the vertex correction $\Lambda\rightarrow\delta_{\sigma ,\sigma'}$.\\

Fig. 4 shows RIXS spectra computed within the 2-band AFM model in
the form of a color plot for momentum transfer along as well as
perpendicular to the direction of the ladders. The d-band spectra
have not been broadened in order to emphasize the presence of
considerable intrinsic structure in the spectra, despite the large
broadening $\Gamma$ associated with the short core hole lifetime.
Insight into the nature of these spectra can be obtained by
examining expressions 1-4 on which the computations are based. The
enhancement factor $W(\omega_i,\omega_f)$ is found to vary
relatively slowly with energy due to the large $4p$ bandwidth and
the substantial damping of the core hole given by $\Gamma_{1s}$.
Therefore, spectral shapes are controlled effectively by the term
$\sum_{k}\delta(\omega+E_{j}(\textbf{k})-E_{j^\prime}(\textbf{k}+\textbf{q}))\times
n_j(\textbf{k})[1-n_{j^{\prime}}(\textbf{q}+\textbf{k})] \mid
X^j_{l\sigma}(\omega,\textbf{k})
X^{j^{\prime}}_{l\sigma^{\prime}}(\omega,\textbf{k+q})\mid^2$. Note
that this involves not only the joint density of states (JDOS)
factor,
$n_j(\textbf{k})[1-n_{j^{\prime}}(\textbf{q}+\textbf{k})]\delta(\omega+E_{j}(\textbf{k})-E_{j^\prime}(\textbf{k}+\textbf{q}))$,
but also the partial electron occupancy of the filled band given by
$\mid X^j_{l\sigma}(\omega,\textbf{k})\mid^2$ and the partial
electron occupancy $ \mid
X^{j^{\prime}}_{l\sigma^{\prime}}(\omega,\textbf{k+q})\mid^2$ of the
empty band. A large contribution thus results when JDOS connects
band extrema, leading to resonant peaks in the RIXS
cross-section.\\
Fig. 5 considers the spectrum at $q_z=\pi/{2c}$ in greater detail
(i.e. corresponding to the vertical yellow line in Fig. 4(a)). The
blue curve in Fig. 5(b) gives the total RIXS cross-section, which of
course involves contributions from all allowed transitions from
either of the two unfilled bands to one of the two empty bands in
the AFM band structure of Fig. 3(b). The red curve in Fig. 5(b)
gives the partial contribution to the spectrum from just the pair of
bands shown in Fig. 5(a), i.e. the lowest occupied and the highest
unoccupied band. In particular, peak C around 3.5 eV in Fig. 5(b)
arises from transitions in (a) marked by the green arrow, while peak
B has its origin in the transitions given by the orange arrow. [Note
that the horizontal shift in the direction of the arrows in Fig.
5(a) is the momentum transfer vector, while the vertical
displacement is the energy transferred in the scattering process.]
Other spectral details can be analyzed in a similar manner and
associated with specific transitions by examining the partial
contributions from various pairs of bands. In particular, the
resonant peak A around 2.5 eV in Fig. 5(b) results from transitions
between the uppermost filled band and the lowest empty band in Fig.
3(a). Along these lines, the intense feature around 3 eV in the
$q_\perp$ spectra of Fig. ~4(b) is found to be associated with
transitions between the uppermost filled band and the lowest empty
band (as a function of $q_\perp$).
\begin{figure}[t]
\vspace*{.2in} \hspace*{-.2in} \centering
\includegraphics[width=9cm,height=11cm]{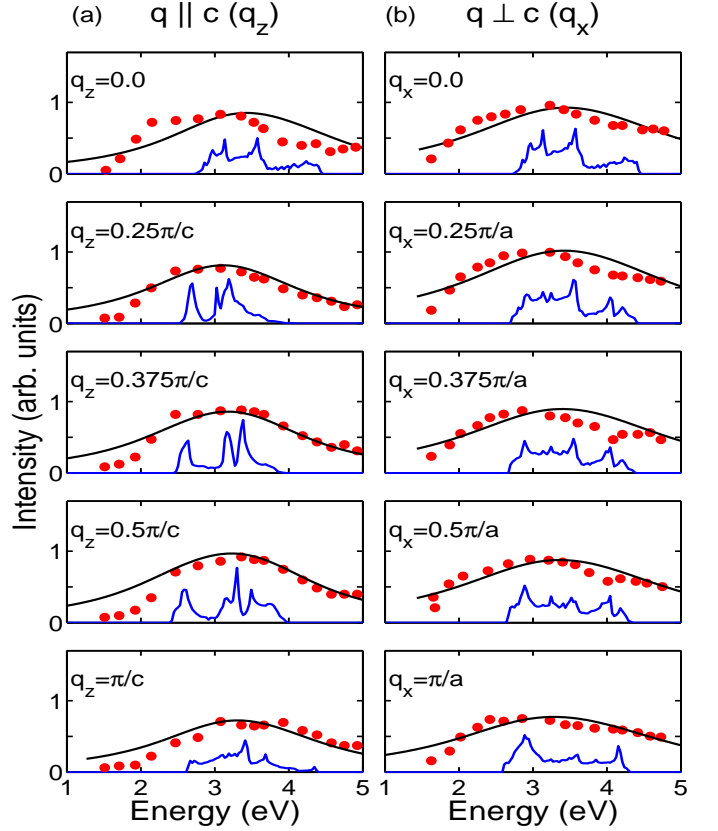}

\caption{(color online) Comparison of unbroadened (blue curves) and
broadened (black curves) RIXS spectra with the corresponding
experimental data (red dots) taken from Ref.~\onlinecite{Hasan}.
Left hand side panels are for $q_\parallel c$ at various $q_z$
values, while the right hand side panels are for $q_\perp c$ at
various $q_x$ values as indicated. Spectra are normalized as shown.}
\label{fig45}
\end{figure}

Fig. 6 compares our theoretical spectra with the available
experimental RIXS data of Ref.~12  on the ladder compound. Left hand
side panels are for momentum transfer along the ladder direction
(i.e. $q_\parallel c$) with $q_z$ varying from 0 to $\pi/c$, while
the right hand side panels are for $q_\perp c$ with $q_x$ varying
over the range 0-$\pi/a$. The unbroadened theoretical spectra (blue
lines) require a substantial broadening for a meaningful comparison
with the data (red dots). Accordingly, we have applied a combined
Gaussian and Lorentzian broadening to the computed spectra to obtain
the broadened theoretical spectra in Fig. 6 (black lines). The
Gaussian broadening is taken as the nominal experimental resolution
of 120 meV\cite{Hasan}. The residual broadening, which reflects
presumably lifetime effects not accounted for in our computations,
is modeled via a Lorentzian with half-width-at-half-maximum of
$\Gamma_d$=1.1 eV for $q\parallel c$ and of $\Gamma_d$=1.4 eV for $q
\perp c$\cite{foot1,ZHRIXS}. Although Figs.~4-6 show that the RIXS
spectra intrisically contain considerable information concerning the
charge excitations and their momentum dependencies in the ladder
compound, much of this structure is seen to be lost in the broadened
spectra. Nevertheless, for both $q\parallel c$ as well as $q\perp
c$, the experimental spectra are in reasonable accord with the
broadened theory, some discrepancies in shape and fine structure in
the experimental data notwithstanding.  In particular, the spectra
show a dispersion of $\sim 0.8eV$ along the ladders ($q_z$) and
negligible dispersion perpendicular to the ladders ($q_x$).  While
we have compared our calculations to the data of Wray {\it et
al.}\cite{Hasan}, the undoped data of Ishii {\it et al.}\cite{Ishii}
show very similar broadening and dispersion, with a slightly larger
gap.
\section{Conclusions}

We have presented a two-band model and the associated K-edge RIXS
spectra for the ladder compound SrCu$_{2}$O$_{3}$ as a way of
capturing the physics of the more complex ladder compound
Sr$_{14}$Cu$_{24}$O$_{41}$. RIXS spectra are considered for momentum
transfer along as well as perpendicular to the direction of the
ladders. Our analysis indicates that the available ARPES and RIXS
data are consistent within experimental resolution with the presence
of strong antiferromagnetic correlations in the system, which are
modeled in our study by introducing an ordered AFM state. Notably,
our calculations do not require any significant renormalization of
LDA-based band paramters in the ladder compound, and moreover, our
value of the effective Hubbard $U$ is similar to that found in the
insulating planar
cuprates\cite{Kusko}.\\
{\bf Acknowledgements}  This work is supported by the U.S.D.O.E
contracts DE-FG02-07ER46352, AC03-76SF00098 and benefited from the
allocation of supercomputer time at NERSC and
Northeastern University's Advanced Scientific Computation Center (ASCC). MZH is supported by DOE/DE-FG02-05ER46200.\\
\section*{Appendix: AFM and Paramagnetic Hamiltonians and Dispersions}

The AFM Hamiltonian for the two band model can be written as
 \be
 \begin{array}{c}
H=\sum_j \varepsilon_0
d_{j}^{+}d_{j}+\sum_{<i,j>}t_{ij}(d_{j}^{+}d_{i}+d_{j}d_{i}^+)+\\\sum_{j}U
n_{d_j\uparrow}n_{d_j\downarrow}
 \end{array}
 \ee
where $n_{d_j}=d_{j}^+d_{j}$, $\varepsilon_0$ is the on site energy,
$t_{ij}$'s are the hopping parameters (see Fig. 1), and $U$ is
Hubbard U. The AFM ordering of spins is shown in Fig. 1 by green
arrows. The Hartree-Fock decomposition of the Hubbard term in the
Hamiltonian is given by
 \be
 \begin{array}{c}
n_{d_j\uparrow}n_{d_j\downarrow}\rightarrow\\
n_{d_j\uparrow}<n_{d_j\downarrow}>+<n_{d_j\uparrow}>n_{d_j\downarrow}-\\<n_{d_j\uparrow}><n_{d_j\downarrow}>\\
U_m=\frac{U}{2}(<n_{d_j\uparrow}>-<n_{d_j\downarrow}>)=Um \\
<n>=<n_{d_j\uparrow}>+<n_{d_j\downarrow}>
 \end{array}
 \ee
 We can now rewrite the AFM Hamiltonian as

 \begin{widetext}
 \begin{eqnarray}
 {\mathcal{H}}_{11}&=&{\mathcal{H}}_{44}= \triangle+U_{m}, ~~~~
{\mathcal{H}}_{22}={\mathcal{H}}_{33}=
\triangle-U_{m}\\\newline\nonumber
 {\mathcal{H}}_{12} &=&
-t_{2}-t_{3}\exp(ik_{x}/2)-t_{6}\exp(3ik_{z}/2)-2t_{8}\cos(2k_{z}),\\\newline\nonumber
 {\mathcal{H}}_{13}
&=&-2t_{1}\cos(k_{z})-2t_{4}\cos((k_{z}-k_{x})/2)-2t_{9}\cos((3k_{z}-k_{x})/2),\\\newline\nonumber
 {\mathcal{H}}_{14}
&=&-2t_{5}\cos(k_{z})-t_{3}\exp(-i(k_{z}-k_{x})/2)-t_{6}\exp(-i(3k_{z}+k_{x})/2),\\\newline\nonumber
 {\mathcal{H}}_{23}
&=&-2t_{5}\cos(k_{z})-t_{3}\exp(i(k_{z}-k_{x})/2)-t_{6}\exp(i(3k_{z}-k_{x})/2),\\\newline\nonumber
 {\mathcal{H}}_{24}
&=&-t_{1}\cos(k_{z})-2t_{4}\cos((k_{z}+k_{x})/2)-2t_{9}\cos((3k_{z}+k_{x})/2)),\\\newline\nonumber
 {\mathcal{H}}_{34}&=&{\mathcal{H}}_{12},\\\newline\nonumber
\end{eqnarray}
where
\begin{eqnarray}
 \triangle
&=&\varepsilon_{\circ}-2t_{7}\cos(2k_{z})-2t_{4}\cos((k_{z}+k_{x})/2)-2t_{9}\cos((3k_{z}-k_{x})/2))\nonumber.
 \end{eqnarray}
 For the the paramagnetic case $U_m$ is equal to zero and the
$4\times4$ AFM Hamiltonian is reduced to a $2\times2$ form with
matrix elements
\begin{eqnarray}
{H}_{11}={H}_{22}&=&\epsilon_0-2t_1\cos(k_z)-2t_7\cos(2k_z)-4t_4\cos(k_x/2)\cos(k_z/2)-4t_9\cos(k_x/2)\cos(3k_z/2),\\\newline\nonumber
{H}_{12}&=&-t_2-2t_5\cos(k_z)-2t_8\cos(2k_z)-2t_3\exp(ik_x/2)\cos(k_z/2)-2t_6\exp(ik_x/2)\cos(3k_z/2),\\\newline\nonumber
\end{eqnarray}
The resulting dispersion is\cite{muller}
 \begin{eqnarray}
\varepsilon_\pm(\textbf{k})=\epsilon_0+\epsilon_\|(k_z)\cos(k_x/2)+(\epsilon_{\bot,1}(k_z)\cos(k+x/2)\pm(\epsilon_{\bot,3}(k_z)^2+\epsilon_{\bot,4}(k_z)^2+2\epsilon_{\bot,3}(k_z)\epsilon_{\bot,3}(k_z)\cos(k_x/2))^{1/2}
 \end{eqnarray}
where
\begin{eqnarray}
\epsilon_\|(k_z)&=&
-2t_1\cos(k_z)-2t_7\cos(2k_z),\\\newline\nonumber
\epsilon_{\bot,1}(k_z)&=&-4t_4\cos(k_z/2)-4t_9\cos(3k_z/2),\\\newline\nonumber
\epsilon_{\bot,3}(k_z)&=&t_2+2t_5\cos(k_z)+2t_8\cos(2k_z),\\\newline\nonumber
\epsilon_{\bot,4}(k_z)&=&
2t_3\cos(k_z/2)+2t_6\cos(3k_z/2),\\\newline\nonumber
\end{eqnarray}

 \end{widetext}


\begin{thebibliography}{999}
\bibitem{foot1}Interestingly, these values of $\Gamma_d$ are
substantially larger than those found previously in analyzing RIXS
data from an electron-doped planar cuprate \cite{ZHRIXS}.
\end{thebibliography}
\end{document}